\documentclass[a4paper,fleqn,usenatbib]{mnras}
\usepackage{newtxtext,newtxmath}
\usepackage[T1]{fontenc}
\usepackage{ae,aecompl}

\usepackage{graphicx}
\usepackage[font=small,skip=0pt]{caption}
\usepackage{deluxetable}
\usepackage{lscape}
\usepackage{epsfig}
\usepackage{amssymb,amsmath}
\newcommand{\ZnII}{\mbox{Zn\,{\sc ii}}}
\newcommand{\AlIII}{\ion{Al}{III}}
\newcommand{\AlII}{\ion{Al}{II}}
\newcommand{\HI}{\ion{H}{I}}
\newcommand{\CrII}{\ion{Cr}{II}}
\newcommand{\SiII}{\ion{Si}{II}}
\newcommand{\MnII}{\ion{Mn}{II}}
\newcommand{\MgII}{\ion{Mg}{II}}
\newcommand{\MgI}{\ion{Mg}{I}}
\newcommand{\FeII}{\ion{Fe}{II}}

\newcommand {\apgt} {\ {\raise-.5ex\hbox{$\buildrel>\over\sim$}}\ }
\newcommand {\aplt} {\ {\raise-.5ex\hbox{$\buildrel<\over\sim$}}\ }

\title[\ZnII-Selected QSO Absorption Line Systems]{Neutral-Gas-Phase Metal Abundances in 
\ZnII-Selected Quasar Absorption Line Systems near Redshift $z=1.2$
\thanks{Based on data obtained from the Sloan Digital Sky Survey and on observations made with the 
Hubble Space Telescope, operated by STScI-AURA for NASA/ESA.}}

\author[E. M. Monier et al.]
{\parbox[t]{\textwidth}{\raggedright 
Eric M. Monier,$^{1}$\thanks{e-mail: emonier@brockport.edu}
David A. Turnshek,$^{2}$
Sandhya M. Rao,$^{2}$
Gendith M. Sardane,$^{2}$
and Daniel Burdette$^{1,3}$
}
\vspace*{8pt}\\
$^{1}$Department of Physics, The College at Brockport, State University of New York, Brockport, NY 14420, USA\\
$^{2}$Department of Physics and Astronomy and PITTsburgh Particle physics, Astrophysics, and Cosmology Center (PITT PACC),\\
 University of Pittsburgh, Pittsburgh, PA 15260, USA\\
$^{3}$Institute for Structure and Nuclear Astrophysics (ISNAP), University of Notre Dame, Notre Dame, IN 46556, USA
}

\date{Accepted XXX. Received YYY; in original form ZZZ}

\pubyear{2018}

\begin{document}
\label{firstpage}
\pagerange{\pageref{firstpage}--\pageref{lastpage}}
\maketitle

\begin{abstract}
We present ten metallicity measurements for quasar absorbers near $z=1.2$ that were selected 
for having unusually significant \ZnII\ absorption in their SDSS spectra. Follow-up UV space 
spectroscopy of the Ly$\alpha$ region shows that all ten have damped Ly$\alpha$ (DLA) absorption,
corresponding to neutral hydrogen column densities in the range $2.4\times 10^{20}$ $\le$ N(\HI) $\le$ 
$2.5\times 10^{21}$ atoms cm$^{-2}$, and indicating that the gas is very optically thick and essentially neutral. 
The sample is a very small subset of systems compiled by searching the University of Pittsburgh 
catalog of $\approx 30,000$ intervening \MgII\ absorption line systems in SDSS quasar spectra up to DR7. 
We started by isolating $\approx 3,000$ that had strong \MgII\ absorption in the redshift interval 
$1.0 < z < 1.5$ in brighter background quasars. Of these, 36 exhibited significant absorption near \ZnII. 
Space UV spectroscopy was then obtained for nine of these (25\% of the total), and a tenth system in a 
fainter quasar was found in the HST archives. The result is a 
representative sample of the highest Zn$^+$ columns of gas within the DLA population.
These \ZnII-selected systems define the upper envelope of DLA metallicities near $z=1.2$. They 
show a tight anti-correlation between N(\HI) and [Zn/H], with the higher metallicity systems clearly 
exhibiting more depletion based on [Cr/Zn] values. The various metal-line measurements (Zn, Cr, Si, Fe, Mn) 
indicate evolved neutral gas with $-0.9 \aplt$ [Zn/H]  $\aplt +0.4$.

\end{abstract}


\section{Introduction}

Since the first spectroscopic survey for damped Ly$\alpha$ systems (DLAs) by \citet{WTSC1986}, 
it has been recognized that these galaxy-sized columns of neutral hydrogen (classically defined 
as having N(\HI) $\ge 2 \times 10^{20}$ atoms cm$^{-2}$) can be used to trace the neutral gas 
content of the Universe back to $z\approx5$. Indeed, the results indicate 
that DLAs contain the bulk of the known \HI\ gas mass in the Universe \citep[e.g. see][and references therein]{Raoetal2017,
Noterdaemeetal2012}. Given our ability to identify this gas, an important 
follow-up challenge is to understand its metallicity and dust content. 

Metallicity studies of DLAs are often done using the elements Zn, Cr, Fe, Mn, and Si, which are expected to exist primarily in a singly 
ionized state. The equivalent widths of the weaker, unsaturated transitions of these lines can be directly converted into 
column densities because they lie on the linear part of the curve of growth. \ZnII\ in particular is often taken to be
a direct measurement of the 
metal abundance of the gas, because it follows Fe and is generally not depleted onto dust grains. Cr, on the other hand, is 
readily depleted onto dust grains, so the neighboring \CrII\ lines can be combined with the \ZnII\ measurements to trace
dust depletion in the gas via [Cr/Zn] in moderate- to high-redshift DLAs; this has been realized for some time
(e.g., \citealt*{PBH1990}, \citealt{Pettini1994}).

One would expect the cosmic neutral gas to become more metal-enriched with increasing cosmic time, forming molecular gas and then stars 
as it cycles through galaxies, and this qualitative expectation has been confirmed \citep[e.g. see][and references therein]{Quiretetal2016}.  
However, we do not yet know how metal-enriched DLAs can become. This study investigates the upper limit of metal abundance in DLAs at $z\approx1.2$.
Our aim is to identify and study absorption-line systems for which unusually significant \ZnII\ is detectable 
among the large number of available moderate-resolution Sloan Digital Sky Survey (SDSS) optical quasar spectra.   
Using the volatile element Zn, we can identify the highest columns of metals in the Universe and explore their environments, 
including their associated neutral hydrogen column densities, and the degree to which some refractory elements 
are depleted onto grains. These systems can reveal important information about the upper envelope 
of the N(HI) versus metallicity relation, which otherwise has an intrinsic spread greater than two orders of magnitude (Quiret et al 2016). 

As shown by \citet{RT2000} and \citet*{RTN2006}, DLAs at redshifts $z<1.6$ 
are a subset of strong \MgII\ absorption systems, which are easily identified in Sloan Digital Sky Survey (SDSS) 
quasar spectra \citep{Quideretal2011}. We assume that \ZnII\ systems detectable in SDSS spectra must be a 
subset of strong \MgII\ systems, and this assumption is confirmed by the results of our analysis. However,
UV space spectroscopy is needed to determine if a particular system at these redshifts has a large-enough N(\HI) value to 
be classified as a neutral-gas DLA system.
%or alternatively a weaker subDLA with $1 \times 10^{19} \leq$ N(\HI) $< 2 \times 10^{20}$ atoms cm$^{-2}$. 
Spectroscopy of unsaturated absorption (e.g. \ZnII$\lambda\lambda2026,2062$)\footnote{Saturated absorption 
like \MgII$\lambda2796,2803$ cannot be used to determine metallicity.}, combined with a measurement of N(\HI),
then allows metallicities to be determined.  

Thus, in this contribution we report on a search for \ZnII\ absorption in the spectra of reasonably 
bright SDSS quasars, followed by UV space spectroscopy of Ly$\alpha$ to measure N(\HI) values and 
hence metallicities. We observed Ly$\alpha$ for nine systems with $1.0 \aplt z \aplt 1.4$, which is 25\% 
of the \ZnII\ sample identified in our search of SDSS spectra. We also found a tenth and similar \ZnII\ 
system with an HST archival spectrum in a fainter SDSS quasar.
%which was not part of our searched sample. 
Given the likely range of metallicities, we reasoned that these \ZnII-selected systems might fall in the 
DLA regime, and this turned out to be the case. These ten systems represent a good sample for future follow-up, 
and we use them here to explore the upper envelope of the distribution of neutral-gas-phase metallicities, 
and corresponding metal depletion measures, as a function of N(\HI) near redshift $z = 1.2$. 

In Section \ref{sec:sample} we define the sample and in Section \ref{sec:observations} we present the observations and measurements. The analysis and results 
are given in Section \ref{sec:results}, and in Section \ref{sec:expandingsdss} we briefly consider the potential of expanding the sample.  A discussion and conclusions 
are presented in Section \ref{sec:discussion}.  

\section{Sample}
\label{sec:sample}

We initially considered a sample of nearly 30,000 SDSS quasar \MgII$\lambda\lambda2796,2803$ absorbers, 
with the aim of identifying a subset for which we could obtain UV spectroscopy of Ly$\alpha$. These 
were the \MgII\ absorbers in the University of Pittsburgh catalog of \citet{Quideretal2011}, which originally 
covered up to DR4 ($\approx 16,600$ systems), and was later augmented with additional systems up to DR7 ($\approx 
13,000$ systems). We then isolated those with $1.0 \le z_{abs} \le 1.5$, SDSS $g$ fiber magnitude $<19.1$,\footnote{This 
approximately corresponds to SDSS $g$ model mag $<18.8$.} and \MgII$\lambda2796$ rest equivalent width 
(REW) $W_0^{\lambda2796} \ge 0.5$ \AA. Strong, redshifted \ZnII$\lambda\lambda2026,2062$ absorption 
would be visible in SDSS spectra within these criteria,\footnote{Our search did not require the presence 
of \CrII\ near this region since Cr might be depleted onto dust grains.} would be bright enough to 
make them amenable to space UV spectroscopy near Ly$\alpha$ absorption with a reasonably short integration time, and 
would have redshifts allowing follow-up galaxy identification studies.
As shown by \citet{RTN2006}, the requirement that $W_0^{\lambda2796} \ge 0.5$ \AA\ 
would find all DLAs with N(\HI) $\ge 2\times10^{20}$ atoms cm$^{-2}$ as well as a subset of subDLAs with 
$2\times10^{20} >$ N(\HI) $\ge 1\times10^{19}$ atoms cm$^{-2}$; such systems are empirically known to 
contain a sufficient amount of neutral gas, making them favorable for the existence of Zn$^+$. These steps 
left us with $\approx 3000$ \MgII\ systems to search for absorption at the predicted locations of \ZnII. 

\begin{figure}
\centerline{
\includegraphics[width=1.\columnwidth,clip,angle=0]{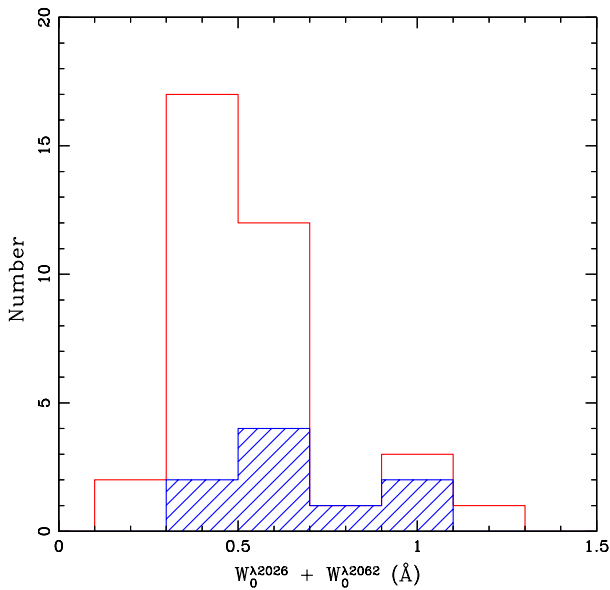}
}
\caption{The red unfilled histogram is the sum of $W_0^{\lambda2026}$ and $W_0^{\lambda2062}$ in 36 
SDSS spectra that meet our \ZnII\ sample inclusion criterion (see Section \ref{sec:sample}).
%Our method of removing weak \MgI\ and \CrII\ from these features allows us to determine the Zn$^+$ column densities (see Sections \ref{sec:observations} 
%and \ref{sec:results}). 
The blue cross-hatched histogram shows the nine for which we obtained N(\HI) from space UV spectroscopy of Ly$\alpha$.
A tenth system with $W_0^{\lambda2026} + W_0^{\lambda2062} \approx 0.4$ \AA\ was found in the HST archives in a fainter quasar.}
% Metallicities were determined for these ten \ZnII-selected systems (Section \S4).}
\label{histogram}
\end{figure}

The final criterion for our sample required absorption features near rest wavelengths of $\lambda2026$ 
and $\lambda2062$. The $\lambda2026$ feature is a blend due to \ZnII\ and much weaker \CrII\ and \MgI. 
The $\lambda2062$ feature is a blend due to \ZnII\ and weaker \CrII. We required $\lambda2026$ and 
$\lambda2062$ to be individually present at levels of significance $>2.5\sigma$, plus together have an 
effective significance of $>4\sigma$, where $\sigma_{eff}^2 = \sigma_{\lambda2026}^2 + \sigma_{\lambda2062}^2$. 
This yielded $36$ \ZnII-selected systems, and we obtained space UV spectroscopy of nine of them (25\%) 
near the predicted position of their Ly$\alpha$ absorption to measure N(\HI). A tenth system 
in a fainter quasar (1017+5356, PI:Becker) found in the HST archives\footnote{Obtained from the Mikulski Archive for
Space Telescopes (MAST).} rounds out the sample. A histogram of the 36 \ZnII-selected systems, and the nine for which 
we obtained space UV spectroscopy of Ly$\alpha$, is shown in Figure \ref{histogram}.   

\section{Observations and Measurements}
\label{sec:observations}

The Journal of Observations for the ten \ZnII-selected systems with space UV spectroscopy is presented in 
Table \ref{tab:Journal_Obs}. In addition to the SDSS coordinates, MJD-plate-fiber IDs, and $g$ fiber magnitudes
of the quasars, Table \ref{tab:Journal_Obs} includes z$_{em}$ and z$_{abs}$ of the systems in our sample. The dates and
spectral elements of space UV spectroscopy, including the signal-to-noise ratio per resolution element in the region of the
Ly$\alpha$ line, are also given. Nine observations were made with HST STIS-G230L ($R\approx 850$), and one was made 
with GALEX NUV ($R\approx 117$). The GALEX observation was previously published in \citet{Monieretal2009}.

\begin{table*}
\centering
\caption{Journal of Space UV Observations}
\label{tab:Journal_Obs}
\begin{tabular}{ccccccccccc}
\hline
\hline
 QSO name & MJD & Plate & Fiber & \it{g} & z$_{em}$ & z$_{abs}$ & Date & Mode & Exp Time & SNR  \\
 & & & & & & & & & [sec] \\

\hline
J0203$-$0910   & 52149 & 666 & 109 & 18.64 & 1.580 & 1.028 & Nov 02 2007  & GALEX NUV      & 30,720 & 14.7 \\
J0254$-$0734   & 51901 & 457 & 495 & 18.08 & 1.472 & 1.288 & Aug 11 2011  & HST STIS-G230L & 2416 & 12.7 \\
J0831+3635     & 52312 & 827 & 001 & 18.57 & 1.160 & 1.127 & May 17 2011  & HST STIS-G230L & 2463 & 11.3 \\
J0953+6351     & 51943 & 487 & 372 & 18.24 & 1.313 & 1.239 & May 12 2011  & HST STIS-G230L & 3686 & 10.6 \\
J1017+5356     & 52381 & 904 & 500 & 19.53 & 1.398 & 1.307 & May 05 2002  & HST STIS-G230L & 5580 & 7.8 \\
J1204+0338     & 52017 & 516 & 589 & 18.72 & 1.545 & 1.208 & Jan 20 2011  & HST STIS-G230L & 2396 & 7.8 \\
J1205$-$3144   & 53474 & 3181 & 576 & 18.74 & 1.689 & 1.274 & Jan 14 2011  & HST STIS-G230L & 2437 & 6.4 \\
J1353+0333     & 52375 & 855 & 096 & 18.93 & 1.584 & 1.208 & Mar 11 2011  & HST STIS-G230L & 5376 & 13.4 \\
J1355+0042     & 51942 & 301 & 321 & 19.09 & 1.624 & 1.404 & Jul 27 2011  & HST STIS-G230L & 5368 & 6.4 \\
J1453+3136     & 55364 & 3875 & 586 & 18.85 & 1.851 & 1.251 & Sep 23 2011  & HST STIS-G230L & 5458 & 4.9  \\
\hline
\end{tabular}
\end{table*}

\subsection{Metal Lines}
\label{sec:SDSS}

As explained in Section \ref{sec:sample}, these quasars were selected on the basis of strong \ZnII\ absorption. The
other available metal lines in the SDSS spectra were also measured (see below) to study the properties of the 
ten systems for which we have UV spectroscopy. For each system, Table \ref{tab:sdssmetaldata} presents the 
rest wavelengths, oscillator strengths \citep{Morton2003}, and rest equivalent widths of the metal lines measured
in the SDSS spectrum.\footnote{The more recent \ZnII\ oscillator strengths 
of \citet{Kisielius2015} are slightly higher than those of \citet{Morton2003} and would lower our 
subsequent determination of [Zn/H] by about 0.1 dex.}

We start by drawing attention to the result that the ten \ZnII\ systems have $1.8 \aplt W_0^{\lambda2796} 
\aplt 3.0$ \AA, which is significantly isolated to higher values than our sample search criteria of 
$W_0^{\lambda2796} > 0.5$ \AA. The total sample of 36 \ZnII\ systems has $0.8 \aplt W_0^{\lambda2796} 
\aplt 4.1$ \AA. Thus, the \MgII$\lambda2796$ properties of the ones selected for space UV spectroscopy 
are similar to the ones not selected, but with a narrower range of values. The very high values for 
$W_0^{\lambda2796}$ generally confirm our assumption that \ZnII\ systems detectable in SDSS spectra are a 
subset of strong \MgII\ systems, and that they may be a subset of the strongest \MgII\ systems.

\begin{table*}
\begin{scriptsize}
\caption{Rest Equivalent Widths (\AA) of Ions Measured in the Sample from SDSS spectra\label{tab:sdssmetaldata}}
\begin{tabular}{cccccccccccccccccc}
\hline
\hline
 & & & \multicolumn{2}{c}{SDSS J0203$-$0910} & \multicolumn{2}{c}{SDSS J0254$-$0734} & \multicolumn{2}{c}{SDSS J0831+3635}
 & \multicolumn{2}{c}{SDSS J0953+6351}  &
\multicolumn{2}{c}{SDSS J1017+5356} \\
Ion & $\lambda_{rest}$ & $f$ & \multicolumn{2}{c}{$z_{abs}$=1.028} & \multicolumn{2}{c}{$z_{abs}$=1.288} & \multicolumn{2}{c}{$z_{abs}$=1.127} & \multicolumn{2}{c}{$z_{abs}$=1.239}
& \multicolumn{2}{c}{$z_{abs}$=1.306} \\
%\cline{4.5-4.5} \cline{6.5-6.5} \cline{8-8} \cline{10-10}
 & (\AA) & &  $W_0$\tablenotemark{a} & $\sigma_{W_0}$  & $W_0$\tablenotemark{a} & $\sigma_{W_0}$
&  $W_0$\tablenotemark{a} & $\sigma_{W_0}$  & 
$W_0$\tablenotemark{a} & $\sigma_{W_0}$  &
$W_0$\tablenotemark{a} & $\sigma_{W_0}$  \\

\hline
          \SiII   & 1808.01  & 0.0022 & \nodata\tablenotemark{b}  & \nodata &  0.299  & 0.041   & 0.309   & 0.081   & 0.277  & 0.048 & 0.375 &  0.059 \\
 \ZnII/\CrII/\MgI & 2026     & \nodata & 0.333  & 0.091    &  0.245  & 0.039   & 0.215   & 0.073   & 0.222  & 0.050 & 0.357 &  0.069 \\
           \ZnII  & 2026.14  & 0.501   &(0.298) & \nodata  & (0.225)  & \nodata &(0.188)   &\nodata  &(0.196)&\nodata&(0.347)& \nodata \\
           \CrII  & 2026.27  & 0.001   & \nodata\tablenotemark{c} & \nodata & (0.002) & \nodata & (0.002)   &\nodata &(0.006) &\nodata& \nodata\tablenotemark{c}  & \nodata  \\
           \MgI   & 2026.48  & 0.113   &(0.035) & \nodata  & (0.018)  & \nodata &(0.025)  &\nodata  &(0.020) &\nodata&(0.01)& \nodata  \\
           \CrII  & 2056.25  & 0.103   & 0.235  & 0.050    &  0.093  & 0.035   & 0.044   & 0.074   & 0.130  & 0.044 & 0.074 & 0.070   \\
      \CrII/\ZnII & 2062     & \nodata & 0.267  & 0.075    &  0.174  & 0.042   & 0.296   & 0.067   & 0.311  & 0.050 & 0.144 & 0.098    \\
           \CrII  & 2062.23  & 0.076   &(0.175) & \nodata  &(0.069)  & \nodata &(0.033)  &\nodata  &(0.097) &\nodata&(0.056)& \nodata \\
           \ZnII  & 2062.66  & 0.246   &(0.092) & \nodata  &(0.104)  & \nodata &(0.263)  &\nodata  &(0.213) &\nodata&(0.088)& \nodata \\
           \CrII  & 2066.16  & 0.051   & 0.381  & 0.063    &  0.050  & 0.033   & 0.064   & 0.046   & 0.117  & 0.057 & 0.093 & 0.060   \\
           \FeII  & 2249.88  & 0.002   & 0.371  & 0.053    &  0.158  & 0.050   & \nodata\tablenotemark{d} & \nodata & 0.252 & 0.077 & 0.247 & 0.085   \\
           \FeII  & 2260.78  & 0.002   & 0.396  & 0.044    &  0.252  & 0.054   & 0.260   & 0.074   & 0.241  & 0.041 & 0.093 & 0.046  \\
           \FeII  & 2344.21  & 0.114   & 1.287  & 0.035    &  1.443  & 0.053   & 1.261   & 0.078   & 1.281  & 0.051 & 0.858 & 0.057  \\
           \FeII  & 2374.46  & 0.031   & 1.108  & 0.034    &  0.943  & 0.055   & 0.904   & 0.065   & 0.826  & 0.042 & 0.522 & 0.052  \\
           \FeII  & 2382.77  & 0.320   & 1.791  & 0.033    &  1.865  & 0.060   & 1.606   & 0.072   & 1.410  & 0.047 & 0.836 & 0.052  \\
           \MnII  & 2576.88  & 0.351   & 0.337  & 0.044    &  0.181  & 0.087   & 0.258   & 0.075   & 0.243  & 0.053 & 0.223 & 0.048  \\
           \FeII  & 2586.65  & 0.069   & 1.430  & 0.036    &  1.367  & 0.060   & 1.375   & 0.080   & 1.105  & 0.057 & 0.737 & 0.055  \\
           \MnII  & 2594.50  & 0.280   & 0.335  & 0.058    &  0.143  & 0.049   & 0.225   & 0.064   & 0.200  & 0.054 & 0.083 & 0.062  \\
           \FeII  & 2600.17  & 0.239   & 1.780  & 0.045    &  1.833  & 0.053   & 1.629   & 0.074   & 1.318  & 0.066 & 0.928 & 0.057  \\
           \MnII  & 2606.46  & 0.198   & 0.176  & 0.045    &  0.112  & 0.042   & 0.293   & 0.086   & 0.142  & 0.051 & 0.039 & 0.039  \\
           \MgII  & 2796.35  & 0.612   & 2.665  & 0.056    &  2.510  & 0.043   & 2.619   & 0.061   & 2.343  & 0.041 & 2.841 & 0.077  \\
           \MgII  & 2803.53  & 0.305   & 2.450  & 0.062    &  2.410  & 0.043   & 2.552   & 0.060   & 2.223  & 0.033 & 2.701 & 0.054  \\
           \MgI   & 2852.96  & 1.830   & 0.890  & 0.056    &  0.573  & 0.049   & 0.793   & 0.052   & 0.643  & 0.041 & 0.289 & 0.044  \\
\hline
\vspace{-0.3in}
\end{tabular}
%\end{tiny}
\end{scriptsize}
\tablenotetext{a}{Parentheses denote inferred or estimated values}
\tablenotetext{b}{Not covered by spectrum.}
\tablenotetext{c}{Noise spike}
\tablenotetext{d}{Assumed to be negligible.}
\end{table*}

\begin{table*}
\begin{scriptsize}
\contcaption{Rest Equivalent Widths (\AA) of Ions Measured in the Sample from SDSS Spectra\label{metaldata}}
\begin{tabular}{cccccccccccccccccc}
\hline
\hline
& & & \multicolumn{2}{c}{SDSS J1204+0338}  & \multicolumn{2}{c}{SDSS J1205$+$3144} & \multicolumn{2}{c}{SDSS J1353+0333}
 & \multicolumn{2}{c}{SDSS J1355+0042}  & \multicolumn{2}{c}{SDSS J1453+3136} \\
Ion & $\lambda_{rest}$ & $f$ & \multicolumn{2}{c}{$z_{abs}$=1.208} & \multicolumn{2}{c}{$z_{abs}$=1.273} & \multicolumn{2}{c}{$z_{abs}$=1.208} & \multicolumn{2}{c}{$z_{abs}$=1.404}
& \multicolumn{2}{c}{$z_{abs}$=1.250} \\
%\cline{5-5} \cline{8-8} \cline{11-11} \cline{14-14}
 & (\AA) & &  $W_0$\tablenotemark{a} & $\sigma_{W_0}$  & $W_0$\tablenotemark{a} & $\sigma_{W_0}$  &
$W_0$\tablenotemark{a} & $\sigma_{W_0}$
&  $W_0$\tablenotemark{a} & $\sigma_{W_0}$  & $W_0$\tablenotemark{a} & $\sigma_{W_0}$  \\
\hline
         \SiII   & 1808.01  & 0.0022  & 0.406  & 0.071 & 0.481  & 0.046 & 0.341   & 0.053   & 0.491  & 0.068   & 0.482    & 0.057   \\
 \ZnII/\CrII/\MgI & 2026     & \nodata & 0.279  & 0.088 & 0.465  & 0.068 & 0.367   & 0.108   & 0.531  & 0.058   & 0.271    & 0.040   \\
           \ZnII  & 2026.14  & 0.489   &(0.243) &\nodata&(0.410) &\nodata&(0.333)  &\nodata  &(0.500) &\nodata  &(0.232)   &\nodata   \\
           \CrII  & 2026.27  & 0.005   &(0.005) &\nodata&(0.018) &\nodata&(0.004)  &\nodata  &(0.002) &\nodata  &(0.018)   &\nodata  \\
           \MgI   & 2026.48  & 0.112   &(0.031) &\nodata&(0.037) &\nodata&(0.030)  &\nodata  &(0.029) &\nodata  &(0.020)   &\nodata  \\
           \CrII  & 2056.25  & 0.105   & 0.112  & 0.064 & 0.391  & 0.040 & 0.082   & 0.091   & 0.334  & 0.064   & 0.217    & 0.040   \\
      \CrII/\ZnII & 2062     & \nodata & 0.271  & 0.072 & 0.473  & 0.052 & 0.360   & 0.134   & 0.319  & 0.054   & 0.289    & 0.053   \\
           \CrII  & 2062.23  & 0.078   &(0.084) &\nodata&(0.293) &\nodata&(0.062)  &\nodata  &(0.212) &\nodata  &(0.163)   &\nodata  \\
           \ZnII  & 2062.66  & 0.256   &(0.187) &\nodata&(0.180) &\nodata&(0.298)  &\nodata  &(0.107) &\nodata  &(0.126)   &\nodata   \\
           \CrII  & 2066.16  & 0.052   & 0.040 & 0.050  & 0.269  & 0.044 & 0.026   & 0.070   & 0.089  & 0.048   & 0.143    & 0.044   \\
           \FeII  & 2249.88  & 0.00182 & 0.133  & 0.068 & 0.457  & 0.038 & \nodata\tablenotemark{d} & \nodata & 0.234  & 0.076   & 0.152    & 0.039   \\
           \FeII  & 2260.78  & 0.00244 & 0.255  & 0.082 & 0.637  & 0.036 & 0.250   & 0.072   & 0.422  & 0.098   & 0.280    & 0.039   \\
           \FeII  & 2344.21  & 0.114   & 2.067  & 0.082 & 2.164  & 0.047 & 1.347   & 0.104   & 1.694  & 0.087   & 1.858    & 0.036   \\
           \FeII  & 2374.46  & 0.0313  & 1.366  & 0.084 & 1.928  & 0.048 & 1.000   & 0.088   & 1.396  & 0.093   & 1.177    & 0.034   \\
           \FeII  & 2382.77  & 0.320   & 2.329  & 0.088 & 2.391  & 0.050 & 1.412   & 0.093   & 1.973  & 0.088   & 2.291    & 0.039   \\
           \MnII  & 2576.88  & 0.3508  & 0.469  & 0.109 & 0.645  & 0.060 & 0.222   & 0.134   & 0.403  & 0.075   & 0.328    & 0.049   \\
           \FeII  & 2586.65  & 0.0691  & 1.951  & 0.099 & 2.263  & 0.067 & 1.382   & 0.108   & 1.654  & 0.080   & 1.640    & 0.042   \\
           \MnII  & 2594.50  & 0.271   & 0.362  & 0.170 & 0.885  & 0.105 & 0.319   & 0.114   & 0.477  & 0.103   & 0.254    & 0.049   \\
           \FeII  & 2600.17  & 0.239   & 2.483  & 0.106 & 2.573  & 0.069 & 1.547   & 0.107   & 1.950  & 0.085   & 1.928    & 0.061   \\
           \MnII  & 2606.46  & 0.1927  & 0.258  & 0.120 & 0.395  & 0.084 & 0.273   & 0.102   & 0.173  & 0.069   & 0.213    & 0.050   \\
           \MgII  & 2796.35  & 0.6123  & 2.628  & 0.064 & 3.036  & 0.064 & 1.813   & 0.070   & 2.657  & 0.059   & 2.787    & 0.065   \\
           \MgII  & 2803.53  & 0.3054  & 2.446  & 0.060 & 2.794  & 0.056 & 1.817   & 0.073   & 2.535  & 0.052   & 2.168    & 0.062   \\
           \MgI   & 2852.96  & 1.810   & 0.980  & 0.103 & 1.193  & 0.048 & 0.962   & 0.125   & 0.926  & 0.073   & 0.653    & 0.049  \\
\hline
\vspace{-0.3in}
\end{tabular}
\tablenotetext{a}{Parentheses denote inferred or estimated values}
\tablenotetext{d}{Assumed to be negligible.}
\end{scriptsize}
\end{table*}

To make the measurements presented in Table \ref{tab:sdssmetaldata}, the SDSS spectra were normalized by continuum fits in the 
standard way \citep[e.g.][]{Quideretal2011}. The results are shown for the visible \SiII, \ZnII, \CrII, \MnII, 
and some \FeII\ transitions in the absorber rest-frame in the middle two and right panels of Figure 
\ref{samplespectra}. The REWs of these lines and others were measured by fitting Gaussians to the absorption 
features. The only problematic aspect is that some absorption features are blends at the SDSS resolution. 
Notably, \ZnII$\lambda2026$ (with oscillator strength $f=0.501$) blends with \MgI $\lambda2026$  ($f=0.113$) 
and \CrII$\lambda2026$ ($f=0.001$), although the \MgI\ contribution is minimal and the \CrII\ contribution 
is generally insignificant. The \MgI$\lambda2026$ contribution can be estimated and removed since we know 
the strength of \MgI $\lambda2852$ ($f=1.830$). Also, \ZnII$\lambda2062$ ($f=0.246$) is blended with 
\CrII$\lambda2062$ ($f=0.076$), whereas \CrII\ lies alone at $\lambda2056$ ($f=0.103$) and $\lambda2066$ 
($f=0.051$). By taking advantage of the fact that these weak lines lie on the linear part of the curve of growth, 
and using the \CrII\ equivalent widths and the oscillator strength ratios, we are able to infer the \ZnII\ 
contribution in the blends as described in \citet{Nestoretal2003}. 
%Also, see Section \S4.
%Typical REWs of significant metal lines (i.e. $>2\sigma$) were in the range of 200-500 m\AA. 

\begin{figure*}
\vspace{0.0in}\centerline{
\includegraphics[width=\textwidth]{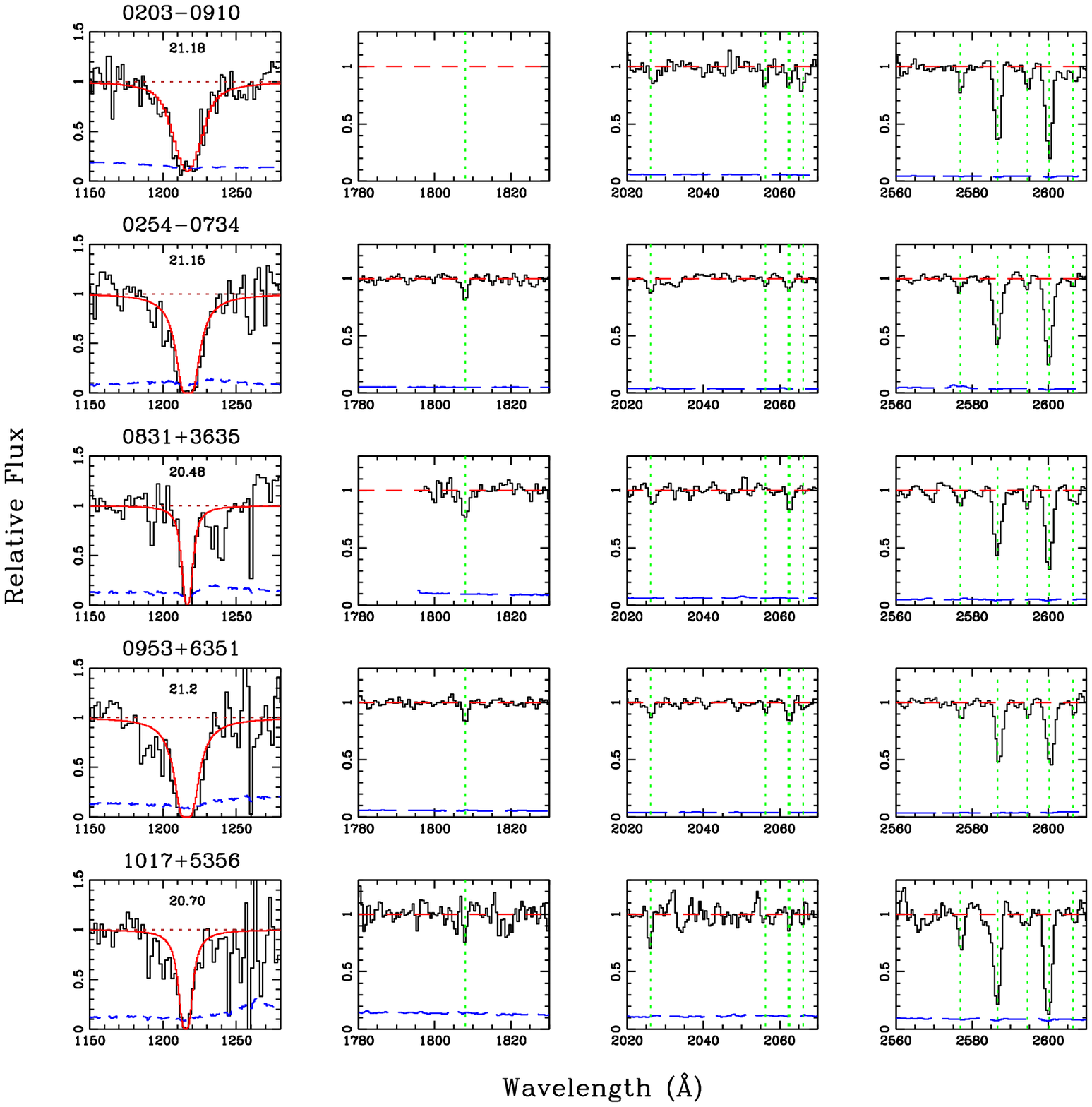}}
\caption{Absorption lines for the first five of the ten systems are shown. From left to right, the four panels are the 
continuum-normalized, rest-frame spectra in the regions containing (1) Ly$\alpha$; (2) \SiII$\lambda1808$; 
(3) \ZnII$\lambda\lambda2026,2062$ and \CrII$\lambda\lambda2056,2066$; and (4) \MnII$\lambda\lambda\lambda2576,2594,2606$ 
and \FeII$\lambda\lambda2586,2600$. The $\lambda2026$ and $\lambda2062$ features are weak blends, as noted in Table \ref{tab:sdssmetaldata}.
Other metal lines in these systems can be readily viewed on the SDSS website. For all spectra, the red dashed line shows the fitted 
continuum and the lower blue dashed line shows the 1$\sigma$ error array. The HST-STIS UV spectra have been binned by three pixels
for presentation. The best-fit DLA profile is shown superimposed in red on the Ly$\alpha$ line, 
and the corresponding value of $\log$ [N(\HI)/cm$^{-2}$] is noted above the fit. For the metal 
lines, the optical SDSS spectra have been smoothed by two pixels 
and a green, vertical dashed line is shown at the expected locations of the metal lines.}
\label{samplespectra}
\end{figure*}

\begin{figure*}
\vspace{0.0in}\centerline{
\includegraphics[width=2.2\columnwidth, angle=0]{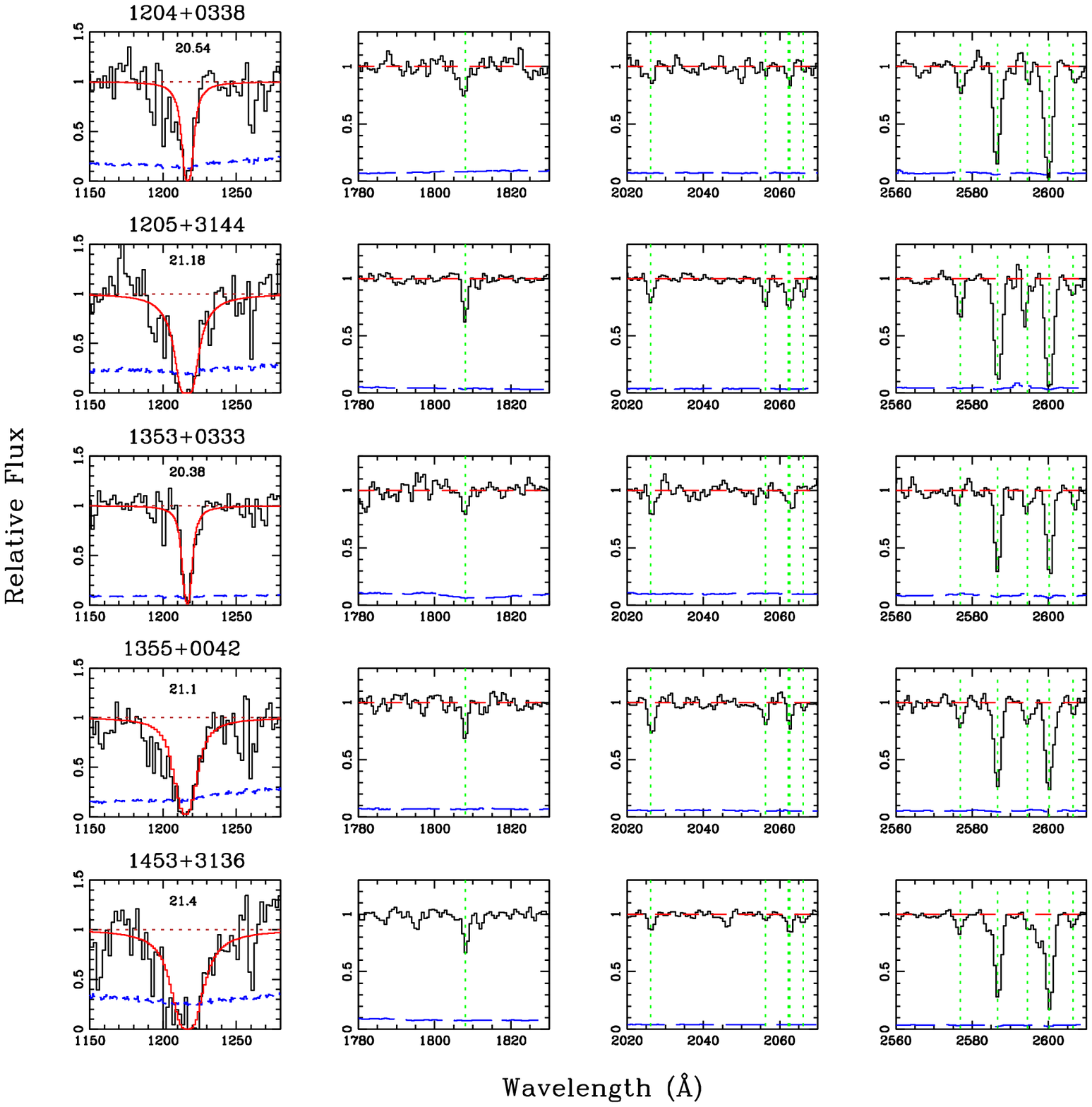}}
\contcaption{HST and SDSS data for the second set of five systems.}
%\label{fig:spectrab}
\end{figure*}

\subsection{Ly$\alpha$ Absorption Lines}
\label{sec:UVspectra}

The aim of the UV spectroscopy (Table \ref{tab:Journal_Obs}) for each \ZnII\ system was simply to measure the Ly$\alpha$ 
absorption line to derive (or place upper limits on) its neutral hydrogen column density, N(\HI). 
Each spectrum was measured by first extrapolating the continuum across Ly$\alpha$ absorption, 
fitting a local continuum, normalizing the spectrum, and then fitting the Ly$\alpha$ with a Voigt 
profile convolved with the instrumental resolution. All of the Voigt profiles showed damping wings, 
allowing us to derive reliable N(\HI) values from the fits. Errors on the N(\HI) values were estimated 
by re-normalizing the continuua by $1\pm\sigma$, where $\sigma$ is the spectrum error array, and re-fitting 
the Voigt profiles. The resulting rest-frame, normalized spectra for the systems near Ly$\alpha$ absorption 
are shown in the leftmost panels of Figure \ref{samplespectra}. The $\log$ [N(\HI)/cm$^{-2}$]
values and derived errors are presented 
in Table \ref{tab:columndensities}. Note that six of the ten systems have $\log$ [N(\HI)/cm$^{-2}$]$>$21.

\subsection{Ancillary Keck Spectroscopy}
\label{sec:keckspectra}

\begin{figure}
\centerline{
\includegraphics[width=1.0\columnwidth,angle=0]{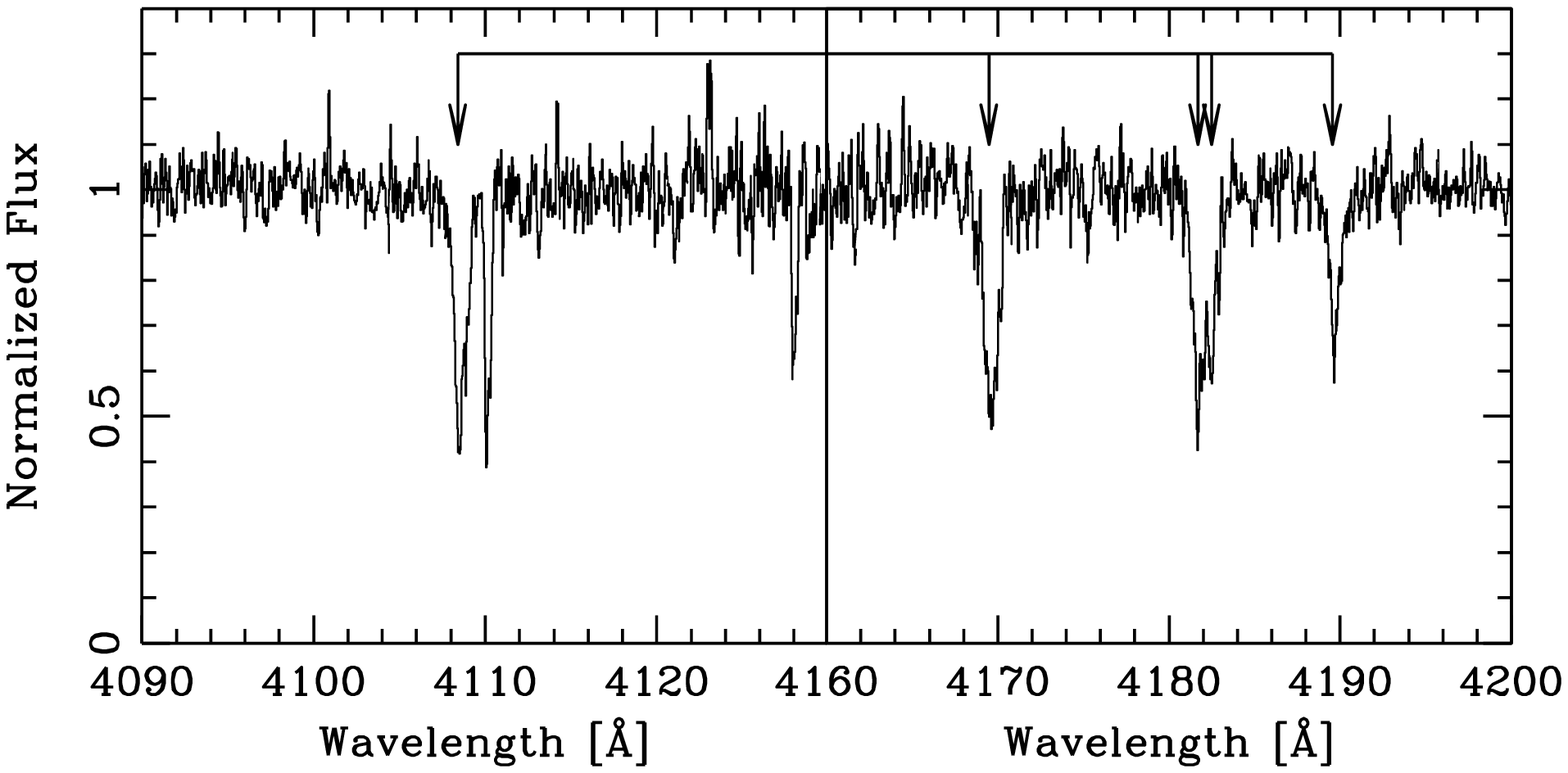}
}
\caption{The \ZnII\ and \CrII\ region of a KOA HIRES spectrum
of 0203$-$0910, smoothed over three pixels. The arrows label the most prominent 
absorptions due to \ZnII$\lambda2026$, \CrII$\lambda2056$, \CrII$\lambda2062$, \ZnII$\lambda2062$, 
and \CrII$\lambda2066$. The wavelength locations of the very weak transitions due to \CrII$\lambda2026$ 
and \MgI$\lambda2026$, which could blend with \ZnII$\lambda2026$, are not labeled and do not appear to be significant. 
Measurements of the equivalent widths of these features generally confirm the results of SDSS 
spectral measurements presented in Table \ref{tab:sdssmetaldata} (see Section \ref{sec:results}).  
Note that the features at the observed wavelengths of $\approx 4110$ \AA\ and $\approx 
4126$ \AA\ are \AlIII$\lambda1854$ and \AlII$\lambda1852$ from another absorption system at $z_{abs} =1.217$ previously
reported in \citet{Monieretal2009}.}
\label{keck0203}
\end{figure}

\begin{figure}
\centerline{
\includegraphics[width=1.0\columnwidth,angle=0]{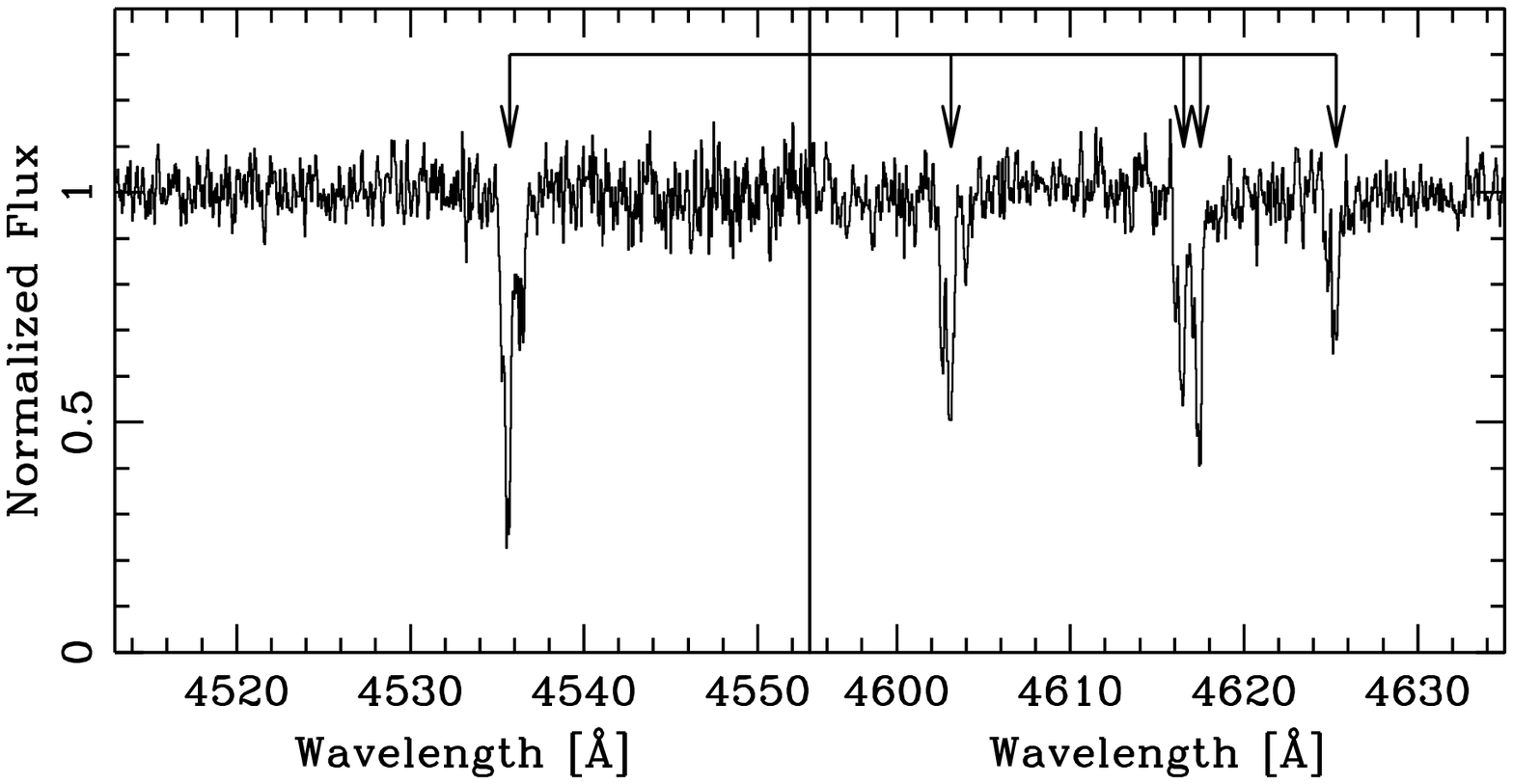}
}
\caption{The \ZnII\ and \CrII\ region of a KOA HIRES spectrum
of 0953+6351, smoothed over three pixels.
The arrows label the most prominent absorptions due to \ZnII$\lambda2026$, \CrII$\lambda2056$, \CrII$\lambda2062$, 
\ZnII$\lambda2062$, and \CrII$\lambda2066$. The wavelength locations of the very weak transitions due to \CrII$\lambda2026$ 
and \MgI$\lambda2026$ are not labeled, but \MgI$\lambda2026$ may be present
at 4536.5 \AA, just to the right of the arrow, with a rest equivalent width of 0.02 \AA.
Measurements of the equivalent widths of these features generally confirm the results of SDSS spectral measurements 
presented in Table \ref{tab:sdssmetaldata} (see Section \ref{sec:results}).}
\label{keck0953}
\end{figure}

A search of the Keck Observatory Archive (KOA) revealed that two of the ten objects in our sample,
0203$-$0910 and 0953$+$6351, also have Keck HIRES spectroscopy. Portions of their Keck spectra 
are shown in Figures \ref{keck0203} and \ref{keck0953}. We discuss these spectra in Section \ref{sec:results}.

\section{Analysis and Results}
\label{sec:results}

The REWs of transitions due to \ZnII, \CrII, \SiII, \MnII, and some \FeII\ lines (Table \ref{tab:sdssmetaldata}) 
are weak and lie on the 
linear part of the curve of growth. They can be converted into column densities using appropriate weighting when 
multiple transitions are observed. These column densities for our \ZnII-selected sample of DLAs are reported in Table
\ref{tab:columndensities}. 

\begin{table*}
\caption{Column Densities [atoms cm$^{-2}$] of Ions\label{tab:columndensities}}
\begin{tabular}{ccccccccccc}
\hline
\hline
 QSO & $\log$ N(\HI)\tablenotemark{a} &  $\log$ N(Zn$^+$)\tablenotemark{a} & $\log$ N(Cr$^+$)\tablenotemark{a} & $\log$ N(Fe$^+$)\tablenotemark{a} & $\log$ N(Mn$^+$)\tablenotemark{a} & $\log$ N(Si$^+$)\tablenotemark{a}  \\
\hline
SDSS J0203$-$0910 & $21.16\pm0.14$ & $13.08\pm0.11$ & $13.87\pm0.06$  & $15.58\pm0.04$ & $13.23\pm0.04$                   & \nodata           \\
SDSS J0254$-$0734 & $21.25\pm0.07$ & $13.08\pm0.07$ & $13.37\pm0.10$  & $15.33\pm0.08$ & $12.94\pm0.10$                   & $15.67\pm0.06$    \\
SDSS J0831+3565   & $20.48\pm0.11$ & $13.15\pm0.10$ & $13.06\pm0.11$  & $15.35\pm0.13$\tablenotemark{b} & $13.15\pm0.08$  & $15.67\pm0.12$    \\
SDSS J0953+6351   & $21.18\pm0.10$ & $13.13\pm0.08$ & $13.54\pm0.08$  & $15.36\pm0.07$ & $13.07\pm0.07$                   & $15.63\pm0.08$    \\
SDSS J1017+5356   & $20.70\pm0.10$ & $13.24\pm0.09$ & $13.31\pm0.24$  & $15.03\pm0.15$ & $12.87\pm0.10$                   & $15.77\pm0.07$    \\
SDSS J1204+0338   & $20.54\pm0.09$ & $13.18\pm0.11$ & $13.45\pm0.13$  & $15.29\pm0.12$ & $13.34\pm0.08$                   & $15.80\pm0.08$    \\
SDSS J1205+3144   & $21.18\pm0.15$ & $13.31\pm0.05$ & $14.03\pm0.03$  & $15.76\pm0.02$ & $13.54\pm0.03$                   & $15.88\pm0.04$    \\
SDSS J1353+0333   & $20.38\pm0.08$ & $13.29\pm0.12$ & $13.29\pm0.19$  & $15.34\pm0.13$\tablenotemark{a} & $13.21\pm0.11$  & $15.73\pm0.07$    \\
SDSS J1355+0042   & $21.08\pm0.13$ & $13.25\pm0.05$ & $13.87\pm0.06$  & $15.52\pm0.08$ & $13.30\pm0.06$                   & $15.89\pm0.06$    \\
SDSS J1453+3136   & $21.40\pm0.22$ & $13.11\pm0.06$ & $13.76\pm0.06$  & $15.36\pm0.05$ & $13.20\pm0.05$                   & $15.88\pm0.05$    \\
\hline
\vspace{-0.3in}
\end{tabular}
\tablenotetext{a}{$\log$ [N(X)/cm$^{-2}$]}
\tablenotetext{b}{\FeII$\lambda2260$ only}
\end{table*}

When DLA metallicity results are reported in the literature, the usual assumption is that most of the metals are 
in the singly ionized state, and therefore significant ionization corrections are not required 
\citep[e.g.][]{Vladilo2001, DZ2003}. This has proven to be a good assumption since the hydrogen is nearly 
completely neutral, and any corrections would generally be below the measurement error. Another often-made 
assumption is that Zn is not significantly depleted onto dust grains. This assumption is reliable at the 
lowest metallicities but is less reliable at high values. It is reported that Zn depletion is $\approx 0.1 - 0.2$ 
dex in most DLAs, but can be up to 0.5 dex for metal-rich DLAs \citep{deCia2018}. However, we will 
not attempt to estimate and correct for any Zn depletion, keeping consistent with the 
way results have been reported in the literature. Thus, the assumption that Zn is not depleted may 
lead to an underestimate of the true Zn metallicity, especially at the highest metallicities.

\begin{table*}
\caption{Metal Abundance Ratios\label{tab:abundances}}
\begin{tabular}{ccccccccccc}
\hline
\hline

 QSO & $\log$ N(HI)\tablenotemark{a} & [Zn/H] & [Cr/H] & [Cr/Zn] & [Mn/H] & [Fe/H] & [Si/H] \\

\hline
SDSS J0203$-$0910  & 21.16 & $-0.64\pm0.18$ & $-0.93\pm0.16$ & $-0.28\pm0.14$ & $-1.36\pm0.15$ & $-1.08\pm0.15$ & \nodata        \\
SDSS J0254$-$0734  & 21.25 & $-0.73\pm0.11$ & $-1.52\pm0.13$ & $-0.78\pm0.14$ & $-1.74\pm0.13$ & $-1.42\pm0.11$ & $-1.09\pm0.10$ \\
SDSS J0831+3565    & 20.48 & $ 0.11\pm0.16$ & $-1.06\pm0.16$ & $-1.17\pm0.16$ & $-0.76\pm0.14$ & $-0.63\pm0.17$ & $-0.32\pm0.16$ \\
SDSS J0953+6351    & 21.18 & $-0.61\pm0.14$ & $-1.28\pm0.13$ & $-0.67\pm0.13$ & $-1.54\pm0.13$ & $-1.32\pm0.13$ & $-1.06\pm0.13$ \\
SDSS J1017+5356    & 20.70 & $-0.02\pm0.14$ & $-1.03\pm0.26$ & $-1.01\pm0.26$ & $-1.26\pm0.15$ & $-1.17\pm0.19$ & $-0.44\pm0.13$ \\
SDSS J1204+0338    & 20.54 & $ 0.08\pm0.15$ & $-0.73\pm0.17$ & $-0.81\pm0.19$ & $-0.63\pm0.13$ & $-0.75\pm0.16$ & $-0.25\pm0.12$  \\
SDSS J1205+3144    & 21.18 & $-0.43\pm0.17$ & $-0.79\pm0.16$ & $-0.37\pm0.09$ & $-1.07\pm0.16$ & $-0.92\pm0.16$ & $-0.81\pm0.16$  \\
SDSS J1353+0333    & 20.38 &  $0.35\pm0.15$ & $-0.73\pm0.21$ & $-1.08\pm0.23$ & $-0.60\pm0.14$ & $-0.54\pm0.16$ & $-0.16\pm0.11$ \\
SDSS J1355+0042    & 21.08 & $-0.39\pm0.15$ & $-0.85\pm0.15$ & $-0.47\pm0.10$ & $-1.21\pm0.14$ & $-1.06\pm0.16$ & $-0.70\pm0.15$ \\
SDSS J1453+3136    & 21.40 & $-0.85\pm0.23$ & $-1.28\pm0.23$ & $-0.43\pm0.11$ & $-1.63\pm0.23$ & $-1.54\pm0.23$ & $-1.03\pm0.23$ \\
\hline
\end{tabular}
\tablenotetext{a}{$\log$ [N(HI)/cm$^{-2}$]}
\end{table*}

Our metallicity results are given in Table \ref{tab:abundances} in the standard way relative to solar values from 
\citet{Asplundetal2009}, [X/H] = $\log$ (N$_X$/N$_H)_{abs} -$ $\log$ (N$_X$/N$_H)_{\odot}$. Table
\ref{tab:abundances} also includes [Cr/Zn], which is a measure of depletion commonly reported in
moderate- to high-redshift DLAs \citep[e.g.,][]{PBH1990,Pettini1994}.

In Section \ref{sec:keckspectra} we noted the existence of Keck HIRES observations for two of our systems, shown
in Figures \ref{keck0203} and \ref{keck0953}. Figure \ref{keck0203} shows the spectrum of 0203$-$0910 obtained during a 60 minute exposure in 2006 (PI: Prochaska, unpublished). 
In these higher-resolution data ($R=36,000$), the $\lambda2062$ blend is resolved into its \ZnII\ and \CrII\ components. In addition,
the data appear to show some velocity structure: relative to the most prominent component, a weaker component of
\ZnII$\lambda2026$ may be present at higher velocity, separated by $\approx +60$ km s$^{-1}$. We assumed two components to the
system and fit the Cr and Zn absorption lines and blends with Voigt profiles to obtain rest equivalent widths. 
The equivalent width measurements of these features generally confirmed - to within 1$\sigma$ - the SDSS results 
reported in Table \ref{tab:sdssmetaldata}.
The corresponding Zn and Cr abundances derived from this spectrum are [Zn/H] $\approx -0.69$ and 
[Cr/H] $\approx -1.09$, within 1$\sigma$ of the values reported in Table \ref{tab:abundances} based on the SDSS data.
The ratio [Cr/Zn] $\approx -0.40$ is within $2\sigma$ of the value in Table \ref{tab:abundances}.

Figure \ref{keck0953} presents the Keck spectrum of 0953+6351 obtained during a 45 minute exposure in 2006 
(PI: Prochaska, unpublished). The $\lambda2062$ blend is again resolved into its \ZnII\ and \CrII\ components. 
This higher-resolution spectrum ($R=36,000$) shows the system comprises a prominent component and at least one weaker component, 
separated from the stronger one by $\approx -30$ km s$^{-1}$. We assumed two components and fit Voigt profiles to the Cr and Zn 
absorption lines and blends. The combined rest equivalent widths of the \ZnII\ and \CrII\ lines
are within 1$\sigma$ of the values determined from the SDSS spectrum (Table \ref{tab:sdssmetaldata}). These measurements of
the Keck spectrum lead to [Zn/H] $\approx -0.76$ and [Cr/H] $\approx -1.23$, within 1$\sigma$ of the values derived 
from the SDSS data and reported in Table \ref{tab:abundances}. 
The ratio [Cr/Zn] $\approx -0.47$ is within $2\sigma$ of the value in Table \ref{tab:abundances}.

\begin{figure}
\centerline{
\includegraphics[width=1.0\columnwidth,clip,angle=0]{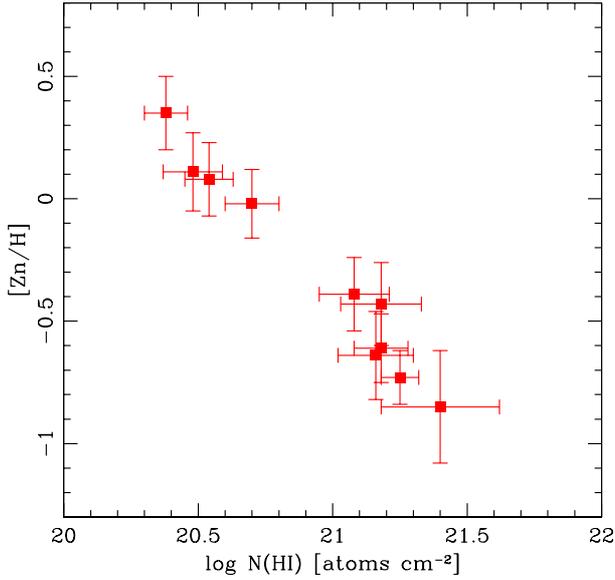}
}
\caption{[Zn/H] as a function of $\log$ [N(\HI)/cm$^{-2}$] (Table \ref{tab:abundances}).}
\label{NHIvsZn}
\end{figure}

\begin{figure}
\centerline{
\includegraphics[width=1.0\columnwidth,clip,angle=0]{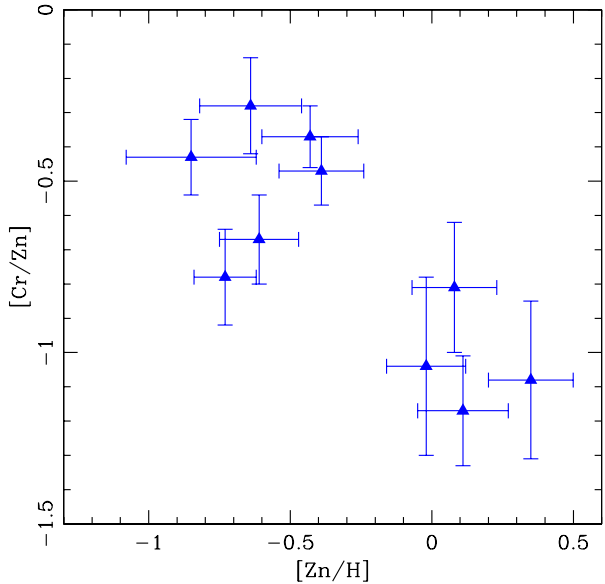}
}
\caption{[Cr/H] as a function of [Zn/H] (Table \ref{tab:abundances}).}
\label{CrvsZn}
\end{figure}

The SDSS spectroscopy shows that the  Zn abundances of the systems in 1353+0333, 0831+3565, 1204+0338 and 1017+5356 are $\approx$ 
224\%, 129\%, 120\% and 95\% of solar, placing these systems among the highest metallicity DLAs yet 
discovered near $z=1.2$. Within our \ZnII-selected sample, these same four systems also have the 
lowest N(\HI) values, with N(\HI) $\aplt 5 \times 20^{20}$ atoms cm$^{-2}$, and the largest depletion 
measures, with [Cr/Zn] $\approx -1$. Figures \ref{NHIvsZn} and \ref{CrvsZn} show these trends by 
plotting [Zn/H] as a function of $\log$ [N(\HI)/cm$^{-2}$] and [Cr/H] as a function of [Zn/H], respectively. 

\begin{figure}
\centerline{
\includegraphics[width=1.0\columnwidth,clip,angle=0]{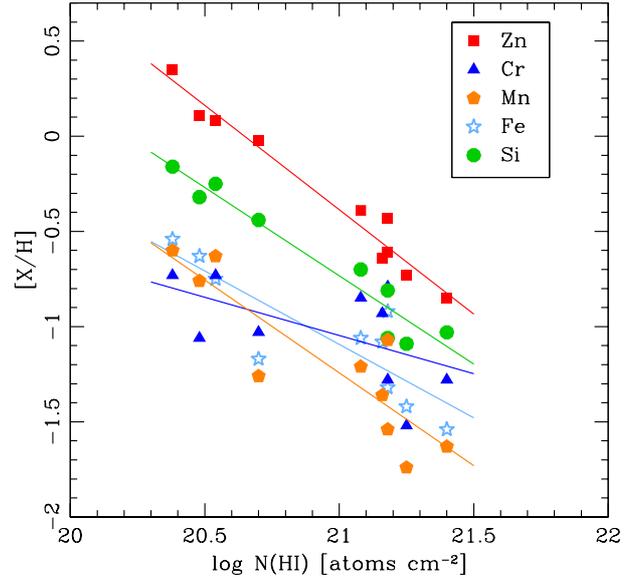}
}
\caption{[X/H] as a function of $\log$ [N(\HI)/cm$^{-2}$] for five metals, corresponding to the color-coded symbols 
in the upper right corner (Table \ref{tab:abundances}). The error bars reported in Table \ref{tab:abundances} have been removed to improve clarity.  }
\label{NHIvsX}
\end{figure}

The metallicities of [Cr/H], [Mn/H], [Fe/H], and [Si/H] are plotted along with [Zn/H] as a function of 
$\log$ [N(\HI)/cm$^{-2}$] in Figure \ref{NHIvsX}, and linear fits are also shown for each data set. 
The error bars reported in Table \ref{tab:abundances} are not shown in Figure \ref{NHIvsX} to make it easier to see trends.

The elements Zn, Fe, and Cr are often assumed to track the Fe-peak elements and to have a common origin. 
Zn itself is not an Fe-peak element but it is produced by the same type of nucleosynthetic processes 
(nuclear statistical equilibrium) that produce normal Fe-peak elements, so it is often argued that it
should generally track them. Therefore, since results indicate that Zn suffers little depletion in DLA
absorbers, past investigators have often argued that it is useful to compare its metallicity to 
Fe-peak elements. At the same time, \citet[and references therein]{Skuladottir2018} have recently
emphasized that although Zn primarily tracks Fe-peak elements in the Milky Way disk and halo stars, it is
not clear that Zn can be used for this purpose in other environments (e.g. in the Milky Way bulge and 
satellite dwarfs).

As shown in Figures \ref{NHIvsZn} and \ref{NHIvsX}, at the highest N(\HI) values where 
metallicity is lowest, [Zn/H] approaches $\approx -0.9$, but at lower N(\HI) where metallicity is largest, 
[Zn/H] approaches $\approx +0.4$. The fit to [Cr/H] (Figure \ref{NHIvsX}) shows a trend of 
increasing depletion of Cr with increasing metallicity (or with decreasing N(\HI) values) relative to 
the [Zn/H] fit, and the fit to [Fe/H] shows a smaller but similar trend.

Our measurements of the Fe-peak element Mn are not atypical of those seen in other DLAs. This is apparent 
if one explores the tabulation for DLAs recently compiled by \citet[and references therein]{Quiretetal2016}.
In our sample, [Mn/Zn]$\approx -0.9$ to within a few tenths of a dex throughout our range of N(\HI) 
values and metallicities. [Mn/Fe] appears to change slightly from $\approx 0.1$ at the lowest N(\HI) 
values and highest metallicities, to values of $\approx -0.3$ at the highest N(\HI) values and lowest 
metallicities. The [Mn/Fe] values are similar to those seen in Galactic halo field stars and individual 
stars in globular clusters for [Fe/H] $\apgt -1.5$ \citep{Sobeck2006}.  The trend we see suggests that 
the origin of the Mn production is core collapse\footnote{Possibly asymmetric.} supernovae prior to the 
time when SN1a are more frequent \citep[][and references therein]{Prantzos2005}; i.e., an increase in [Mn/Fe] is 
expected with increasing metallicity, which is consistent with our measurements. For comparison, the 
work discussed by \citet{SavageandSembach1996} indicates that Fe is significantly more depleted than Mn in 
cool Galactic disk clouds ([Mn/Fe] $\approx +0.8$), but that Mn and Fe are similarly depleted in warm 
Galactic halo clouds ([Mn/Fe] $\approx 0$). Our values of $+0.1\apgt$[Mn/Fe]$\apgt -0.3$ lie mainly outside 
this range. Thus, the Mn measurements offer an important clue to the chemical history of the gas. 
 
The $\alpha$-element Si only shows a minor change relative to Zn as a function of N(\HI) or metallicity. 
We find $-0.5 \aplt$ [Si/Zn] $\aplt -0.2$, with more sub-solar values at lower N(\HI) where [Zn/H] is higher.  
We note that very metal-poor Galactic halo stars, which represent an early stage of chemical evolution, 
show enhanced levels of $\alpha$-elements \citep[e.g.][]{McWilliam1997}, and the lowest-metallicity DLAs
(e.g., [Fe/H]$\aplt -2$, which show little evidence for depletion) often show enhanced levels of $\alpha$-elements
like Si, with [Si/Fe]$\approx +0.3$ \citep[e.g.][and references therein]{Cooke2011}.
None of our \ZnII-selected systems display enhanced Si. Thus, none of the gas in 
these \ZnII-selected systems appears to be in the earliest stages of chemical evolution. This is consistent 
with the result that our \ZnII-selected sample is more chemically evolved with $+0.4\aplt$[Zn/H]$\aplt-0.9$.

\section{Potential of Expanding the Search of SDSS Quasars to Find Additional Detectable \ZnII\ Systems}
\label{sec:expandingsdss}

\begin{figure}
\centerline{
\includegraphics[width=1.0\columnwidth,angle=0]{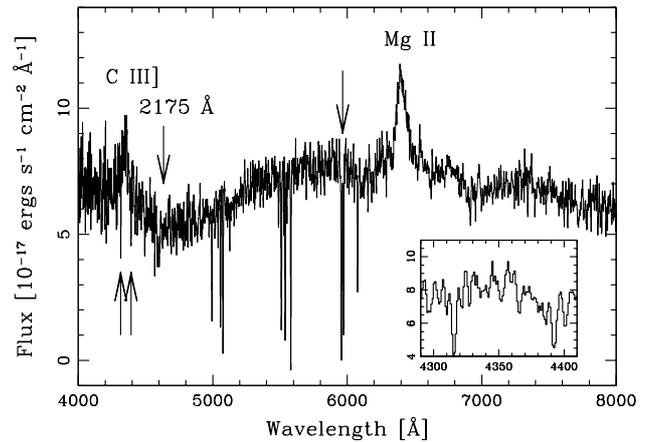}
}
\caption{Observed frame spectrum of SDSS J0831+2138 (z$_{em} = 1.284$). This met our criterion for 
strong \ZnII\ (z$_{abs} = 1.1305$), but was too faint to be included in our \ZnII-selected sample (Section \ref{sec:sample}).  
The \ZnII$\lambda\lambda2026,2062$ absorption features are indicated by the two bottom arrows on the 
lower left; an expanded view is shown in the lower right inset. The \MgII\ absorption doublet near 
6000 \AA\ is labeled with an arrow on the top right. Shortward of the \MgII\ broad emission line the 
f$_{\lambda}$ continuum begins to drop due to 2175 \AA\ dust absorption. }
% and shows the potential of expanding the search of SDSS quasars to find additional detectable \ZnII\ systems. 
\label{Q0831}
\end{figure}

As explained in Section \ref{sec:sample}, of the $\approx 30,000$ SDSS quasar \MgII\ absorption systems we started with, 
after we applied our magnitude and absorption redshift cuts, we ended up with a set of $\approx 3000$ 
to search for strong \ZnII. The cuts were motivated by the practical goal of obtaining a sample that 
could be observed with HST to determine N(\HI) in a reasonable number of orbits. However, it is clear 
that by expanding the sample using a fainter magnitude cut and a broader redshift range, additional 
interesting objects would be found. The tenth \ZnII\ system in a fainter SDSS quasar, which was added 
to our sample because it had HST archival spectroscopy, is one such example. A more interesting example 
is the spectrum of SDSS J0831+2138 shown in Figure \ref{Q0831}. It has a $g$ fiber magnitude of $20.1$, 
so it was 1.0 magnitude too faint to be included in our sample (Section \ref{sec:sample}). There is a clear indication of the 
2175 \AA\ dust absorption feature in its spectrum. However, a discussion of such systems is beyond 
the limited scope of this paper. 

\section{Discussion and Conclusions}
\label{sec:discussion}

\begin{figure}
\centerline{
\includegraphics[width=1.0\columnwidth,clip,angle=0]{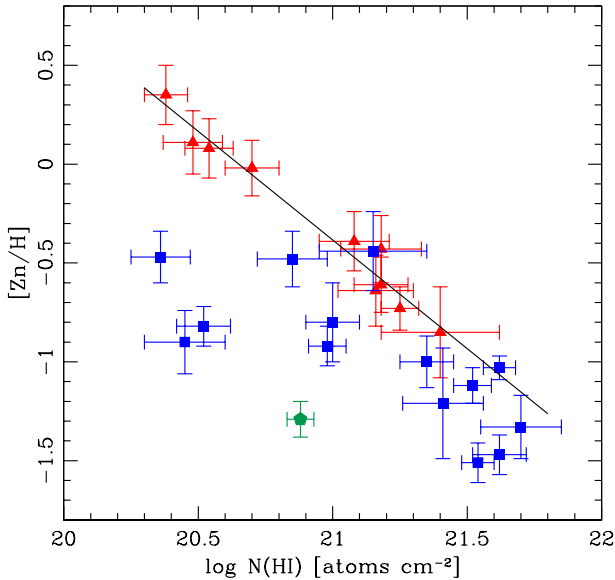}
}
\caption{[Zn/H] values as a function of $\log$ [N(\HI)/cm$^{-2}$] from Table \ref{tab:abundances} for the ten \ZnII-selected systems 
(red triangles), along with values compiled by \citet{Quiretetal2016} from previous work (blue squares). 
The black line represents the metallicity upper envelope as a function of N(\HI), as established by the \ZnII-selected 
sample. The green pentagon is a [Si/H] measurement.}
\label{NHIvsXQuiret}
\end{figure}

The methodology used in this study enabled us to sort through and efficiently identify a sample of
DLAs with the highest Zn$^+$ columns of gas in the redshift interval $1<z<1.5$. We searched
for \ZnII$\lambda\lambda2026,2062$ in $\approx 3000$ relatively bright quasars with known strong \MgII\ 
absorbers, and found that 36 (1.2\%) of them exhibited strong \ZnII\ absorption. UV space spectroscopy of nine of 
these (25\%), plus one in a fainter quasar with HST UV spectroscopy, then allowed us to measure N(\HI) 
values and to establish that they were DLAs. These \ZnII-selected systems have  column densities which 
lie in a relatively narrow range, $13.08 \aplt$ N(Zn$^+$) $\aplt 13.31$, which corresponds to a factor 
of $\approx 1.7$. This narrow range evidently reveals a value for the typical maximum chemical enrichment 
into neutral gas regions that represent the DLAs.  The ten systems studied here only span an N(\HI) range 
of about a factor of ten ($20.4<$ $\log$ [N(\HI)/cm$^{-2}$] $<21.4$). This is a factor of five smaller 
than the observed range of DLA N(\HI) values ($20.3 <$ $\log$ [N(\HI)/cm$^{-2}$] $\aplt 22$). 
Figure \ref{NHIvsXQuiret} is similar to Figure \ref{NHIvsZn}, but in addition to showing 
[Zn/H] as a function of $\log$ [N(\HI)/cm$^{-2}$] for our ten \ZnII-selected systems, it also
shows values for all DLAs with \ZnII\ measurements at $1<z<1.5$ from the compilation of \citet{Quiretetal2016}.
This figure shows how our \ZnII-selected systems establish the upper envelope 
of the $\log$ [N(\HI)/cm$^{-2}$] versus [Zn/H] relation at these redshifts. 

The metallicities of these \ZnII-selected systems ($+0.4\apgt$[Zn/H]$\apgt-0.9$),
along with their corresponding depletion measures, indicate that this 
neutral gas is not representative of gas still in its earliest stage of chemical evolution. But the [Mn/Fe] 
values ($+0.1>$ [Mn/Fe] $>-0.3$) suggest that enrichment happened primarily from core collapse supernovae. 
And while our results trace the upper envelope of the metallicity distribution of DLAs as a function of N(\HI), 
DLAs found in other surveys can fall more than two orders of magnitude 
below this envelope at lower metallicity; i.e., most DLAs are much more metal-poor at a given N(\HI).
In this context, it is interesting that the systems presented here have metallicities
that match the most metal-rich components of halo field stars and globular clusters in external galaxies.
For example, \citet[][and references therein, especially \citealt{Beasleyetal2008} and \citealt{Ibata2014}]{Lamers2017}
discuss these metallicity distributions. Their work indicates
that the halos can possess both metal-poor and metal-rich components; the metal-rich component has a tail that
generally reaches metallicities as high as the DLAs that trace the upper envelope of the metallicity distribution.

A halo origin for the gas is supported by the relatively large impact parameters found by \citet{Rao2011} for
galaxies associated with DLAs at $0.1<z<1.1$.

Greater depletion measures are generally seen when neutral gas is denser \citep[e.g.][and
references therein]{SavageBohlin1979, Spitzer1985, JSS1986}.\footnote{See \citet{Vladilo2002}, \citet{Vladiloetal2011}, 
\citet{JenkinsWallerstein2017}, and references therein for recent work on improved methods to measure depletion 
patterns.} In addition, as dictated by dust condensation temperatures, cooler gas has higher depletion 
measures than warmer gas \citep[e.g.,][see their figure 6]{SavageandSembach1996}. Thus,
our more highly depleted systems, with [Cr/Zn]$\approx$-1, [Zn/H]$\approx$0, and lower N(\HI) 
(Figures \ref{NHIvsZn} and \ref{CrvsZn}),
would represent denser and/or cooler gas, while at the other extreme, 
the systems with [Cr/Zn]$\approx -0.5$, [Zn/H]$\approx -1$, and higher N(\HI) would represent less dense and/or
warmer gas. (See also the different slopes of the fits in Figure \ref{NHIvsX}.) Our overall depletion measures
are broadly consistent with metal-enriched gas residing in halos, with some warm disk gas
mixed in at the highest metallicities. These trends may be
related to the characteristics of the DLA host galaxy environment.

Given these trends, our findings should be considered in the
context of work which has suggested that DLAs obey a mass-metallicity relation \citep[e.g.,][]{Krogageretal2017}, where 
higher-N(\HI) DLAs are preferentially associated with galaxies that have lower mass-luminosity and lower metallicity. 
Relative to this model, our findings suggest that a follow-up DLA galaxy identification study of our sample should reveal 
a positive correlation between metallicity and galaxy luminosity and a negative correlation between N(\HI) and
galaxy luminosity. Since
our sample selection method nicely defines the upper envelope to the negative correlation between
metallicity and N(\HI) (Figure \ref{NHIvsZn}), any correlations found with galaxy luminosity might also be well-defined.

Based on this study, the following conclusions can be reached: 

(1) This sample of DLAs is biased by our \ZnII-selection method. Since they are not randomly selected, 
care should be taken when using them with other samples to draw conclusions. For example, when dealing 
with \MgII-selected samples of DLAs, appropriate care must be taken to infer results on their incidence and 
cosmic mass density \citep{Raoetal2017} and mean metallicity \citep*{RTM2018}. 

(2) At the same time, more detailed follow-up spectroscopic and imaging studies of this sample can 
reveal important information about the nature of the most metal-enriched DLAs and DLA galaxies at $1<z<1.5$. 

(3) The equation which establishes the upper envelope for DLA metallicity at $1<z<1.5$ as a function 
of N(\HI) is [Zn/H] = $1.1\{21.0 -\log[\rm{N}(\HI)/cm^{-2}]\}-0.4$. Since Zn might be depleted by one or two tenths 
of a dex, interpretation of this relationship may require some upward adjustment, especially at high metallicity.  

(4) The measured transitions of other singly ionized refractory elements (Si, Cr, Fe, Mn) provide information 
on the levels of depletion and the extent to which the gas is chemically evolved. 

(5) Taken together, the metallicity and depletion measurements of the gas suggest that it resides in 
halos, consistent with halo-sized impact parameters for DLAs found by \citet{Rao2011}.

\section*{Acknowledgments}

We thank the referee, Max Pettini, for his prompt review and helpful suggestions that improved the presentation of this work.
Support for program \#12308 was provided by NASA through a grant from the Space Telescope Science Institute, 
which is operated by the Association of Universities for Research in Astronomy, Inc., under NASA contract NAS 5-26555. 
D.B. received support from the R.V. Mancuso 
Summer Undergraduate Research Award from the Brockport Physics Department. The SDSS is managed by the Astrophysical 
Research Consortium for the Participating Institutions. The Participating 
Institutions are the American Museum of Natural History, Astrophysical Institute Potsdam, University of Basel, 
University of Cambridge, Case Western Reserve University, University of Chicago, Drexel University, Fermilab, the 
Institute for Advanced Study, the Japan Participation Group, Johns Hopkins University, the Joint Institute for Nuclear 
Astrophysics, the Kavli Institute for Particle Astrophysics and Cosmology, the Korean Scientist Group, the Chinese 
Academy of Sciences (LAMOST), Los Alamos National Laboratory, the Max-Planck-Institute for Astronomy (MPIA), the 
Max-Planck-Institute for Astrophysics (MPA), New Mexico State University, Ohio State University, University of Pittsburgh, 
University of Portsmouth, Princeton University, the United States Naval Observatory, and the University of Washington. 
This research has made use of the Keck Observatory Archive (KOA), which is operated by the W. M. Keck Observatory and the 
NASA Exoplanet Science Institute (NExScI), under contract with NASA.

\bibliographystyle{mnras}
\bibliography{Refs}{}

%\bibitem[Rao, Turnshek, Monier]{RTM2018}
%Rao, S.~M., Turnshek, D.~A., \& Monier, E.~M. 2018, submitted

\label{lastpage}
\end{document}